# Towards post-growth policymaking: Barriers and enablers for wellbeing economy and Doughnut economics initiatives

Laura Angresius[a], Milena Büchs[b], Alessia Greselin[c], Daniel W. O'Neill[a, b]

[a] UB School of Economics, Universitat de Barcelona, C/ de John Maynard Keynes 1-11, 08034 Barcelona, Spain

[b] Sustainability Research Institute, School of Earth, Environment and Sustainability, University of Leeds, Leeds, LS2 9JT, United Kingdom

[c] Faculty of Social Sciences, Tampere University, City centre campus, Main building, Kalevantie 4, 33100 Tampere, Finland

# Abstract

Providing wellbeing for all while safeguarding planetary boundaries requires governments to pursue post-growth policies. An important question is how transformative government-led post-growth initiatives can be within the existing institutional context. While first empirical studies on this question demonstrate some of the current limitations of government-led post-growth initiatives, a broader empirical assessment of underlying barriers and enablers across institutional contexts is so far lacking. To address this gap, we examine wellbeing economy and Doughnut economics government initiatives across governance scales in Europe, New Zealand, and Canada. To identify barriers and enablers as well as priorities for future action, we apply a framework that captures systemic and political dimensions to analyze the data.

Overall, our results suggest that the overarching economic growth paradigm severely limits the initiatives' scope of action. Important enablers of existing initiatives are crises, the political agency of key individuals and high-level political support. Policymakers who promote growth-critical perspectives often face tensions: they need to appeal to broad stakeholder groups while avoiding cooptation. Structural changes and a closer engagement with, and pressure from, civil society are required to support post-growth government initiatives.

**Key words:** Post-growth; transformations; government; wellbeing economy; Doughnut economics

# 1. Introduction

Providing wellbeing for all while staying within planetary boundaries may require moving towards post-growth societies in which the provisioning of human wellbeing and the safeguarding of ecological outcomes are prioritized over economic growth (Jackson, 2021; Slameršak et al., 2024).

We use "post-growth approaches" as an umbrella term to describe different perspectives aiming to foster transformations towards post-growth societies. They all have in common a critique of GDP as a measure of progress, and an emphasis on sustainability, equity, and human wellbeing. Post-growth approaches include degrowth, Doughnut economics, and post-growth-aligned wellbeing economy perspectives (Hayden, 2025; Kallis et al., 2025).

Various actors and strategies could contribute to transforming current economic systems towards post-growth (Barlow et al., 2022). Several scholars suggest that the state can play an important role in this process through post-growth-aligned policy decisions, together with strategies pursued by civil society actors outside of established political institutions (e.g. Hirvilammi, 2020; Koch, 2022a, 2022b; Wright, 2019).

However, existing research suggests that government-led post-growth initiatives have had limited impact so far. For example, McCartney et al. (2023, p. 2) argues that they remain restricted to "window-dressing" activities, and Mason & Büchs (2023, p. 10) find that they adopt "weak post-growth" positions. Existing studies on the role of government-led post-growth initiatives tend to focus on a small number of cases, while a broader empirical assessment of underlying barriers and enablers across institutional contexts is still lacking. To address this gap in the literature and to examine the transformative potential of government-led post-growth initiatives within the current institutional context we ask: which barriers and enablers of post-growth government initiatives are common across wellbeing economy and Doughnut economics initiatives, country contexts, and governance levels? To help answer this question and identify priorities for future actions, we collect primary data from 14 wellbeing economy and Doughnut economics initiatives across 9 European countries, New Zealand, and Canada, and across local, regional, and national governance scales. We also employ an explicit analytical framework which allows us to classify different barriers and enablers.

The term "wellbeing economy" refers to an economy geared towards the provision of human wellbeing within ecological limits (Fioramonti et al., 2022). "Doughnut economics" outlines an "ecologically safe and socially just space for humanity" in which basic human needs are met without transgressing planetary boundaries (Raworth, 2017, p. 45). The two types of initiatives are closely aligned. They are the most prominent examples of government initiatives drawing on post-growth approaches. Thus, in this article we refer to them jointly as "post-growth initiatives". However, it is important to note that the practical implementation of the initiatives may not always fully reflect post-growth principles. While wellbeing economy approaches are adopted on national and regional levels, Doughnut economics initiatives have mostly been adopted by local governments. Thus, studying them together allows us to investigate which barriers and enablers are common across different levels of governance.

To categorize barriers and enablers, we apply what we call a 5Ps framework, consisting of the systemic factors of paradigms (P1) and pressures (P2), and the political dimensions of polity (P3), politics (P4), and policy (P5) (see Section 3 for more details). Applying this framework not only helps with developing a more systematic understanding of key barriers and enablers of government-led post-growth initiatives across institutional contexts but also with identifying priorities for future actions. Overall, our research empirically contributes to understanding the role of government-led wellbeing economy and Doughnut economics initiatives in post-growth transformations and formulates recommendations on how they could be fostered.

This article proceeds as follows: Section 2 reviews previous literature on the role of the state in post-growth transformations and on the barriers and enablers of government-led post-growth initiatives. Section 3 presents our analytical framework in more detail, and Section 4 describes our methods. In Section 5 we present our findings, which we discuss in Section 6. Section 7 concludes.

# 2. Literature review

In this section, we review selected literature on the role of the state in post-growth transformations. Then, we discuss relevant empirical literature on the barriers and enablers of post-growth government initiatives.

## 2.1 The role of the state in post-growth transformations

The literature presents different roles of the state in post-growth transformations depending on the respective theoretical and ideological perspectives taken (Brand, 2016). For example, some authors argue that the state is structurally incapable to overcome its growth-dependencies (Blühdorn, 2020; Hausknost, 2020). However, contributions based on materialist state theories advocate the state as a relevant actor to steer transformations, alongside civil society which operates outside state institutions (Bärnthaler, 2024a; D'Alisa & Kallis, 2020; Koch, 2020b). Koch (2020, p. 127) argues that post-growth transformations will require "a combination of bottom-up mobilisations and action and top-down regulation". According to this perspective, the state is uniquely placed to facilitate and enforce far-reaching changes in societal organization due to its regulatory power (Eckersley, 2021). Bärnthaler (2024a) argues that therefore acquiring top-down agency within the state must form an essential part of strategizing, if degrowth-aligned ideas are to be implemented on a societal scale. In this understanding, internal struggles within the state provide an important terrain of post-growth transformations, and the state is essential to bring about social-ecological changes across society. However, it remains unclear how transformative government-led action can be within the existing institutional framework. Thus, in this article we aim to contribute to this debate by inquiring what we can learn from current government-led post-growth initiatives.

## 2.2 Barriers and enablers for post-growth government initiatives

Here, we focus on reviewing the literature on barriers and enablers for Doughnut economics (Bärnthaler et al., 2024; Hayden & Dasilva, 2022; Mason & Büchs, 2023; Trebeck, 2024; Turrini et al., 2026) and wellbeing economy initiatives (Cattaneo et al., 2025; Khmara & Kronenberg, 2020; Schmid, 2025; Turner & Wills, 2022; Vestergaard & Schmid, 2026). The two types of initiatives are closely aligned and the most prominent examples of government initiatives drawing on post-growth approaches. Their ambition is to systemically shift political priorities and to eventually achieve growth independence.

Since beyond GDP metrics are a common tool used within these initiatives to challenge the focus on economic growth in policymaking, we also cover some empirical literature on the barriers and enablers of beyond GDP initiatives in the sub-sections on barriers and enablers (e.g. Battaglia, 2025; Bleys & Whitby, 2015; Chancel et al., 2014; Wallace, 2019).

### 2.2.1 Barriers

The existing literature identifies a range of barriers to government-led post-growth initiatives, including the dominant economic growth paradigm, lack of political and public support, capacity constraints, as well as path dependencies and entrenched ways of working.

Several authors have emphasized that the economic growth paradigm presents an important barrier for post-growth transformations. The economic growth paradigm characterizes capitalist economies and claims that economic growth is "desirable, imperative, and essentially limitless" (Schmelzer, 2015, p. 264). This paradigm is deeply institutionalized through structural growth dependencies and political priorities (Mason & Büchs, 2023; Schmid, 2025; Turrini et al., 2026). Furthermore, it is culturally entrenched, for instance, through the dominance of neoclassical economics in education and the widespread association of economic growth with wellbeing (Hayden & Dasilva, 2022; Mason & Büchs, 2023).

Existing studies also find that post-growth initiatives lack broad-based political and public support, facing powerful opposing interests in their implementation, also within government (Trebeck, 2024; Vestergaard & Schmid, 2026). For instance, Cattaneo et al. (2025) and Turner & Wills (2022) find that Doughnut economics initiatives sometimes prioritize political feasibility over scientific recommendations. Even in the six Wellbeing Economy Governments (see Section 4.1), post-growth proponents remain a minority within public administrations (Mason & Büchs, 2023). Economic actors and vested interests also pose significant resistance to wellbeing economy initiatives more broadly (Bärnthaler et al., 2024; Mason & Büchs, 2023).

Capacity constraints, especially at the local level, have also been identified as a barrier as they limit the ability to implement changes in a multi-level governance context (Trebeck, 2024; Turner & Wills, 2022). Finally, several authors state that path dependencies and entrenched ways of working in political institutions restrict transformative policy changes, for example through short-termism and siloed organization (Cattaneo et al., 2025; Turner & Wills, 2022; Turrini et al., 2026).

The beyond GDP literature also identifies several of the factors found in the broader literature on government-led post-growth initiatives, for instance siloed ways of working, institutional resistance to change and a lack of convincing narratives (e.g. Bache, 2020; Battaglia, 2025; Bleys & Whitby, 2015; Whitby, 2014).

### 2.2.2 Enablers

Post-growth government initiatives and beyond GDP initiatives share many enablers, such as leadership from individuals, coherence across governance scales and knowledge (e.g. Battaglia, 2025; Bleys & Whitby, 2015; Chancel et al., 2014; Wallace, 2019). However, while in the post-growth literature crises are mainly discussed as enabling far-reaching changes (e.g. Buch-Hansen, 2018), empirical studies on beyond GDP metrices find crises to hold a conflicting role (Whitby, 2014). While Bache (2020) finds that the financial crises led to the rise of wellbeing on the government agenda in the UK, Battaglia (2025) and Bleys and Whitby (2015) found that in moments of crises governments clinged to traditional economic views and instruments rather than being open to alternative metrics.

Existing studies find that collaboration with civil society (Trebeck, 2024; Turrini et al., 2026), leadership from within administration (Schmid, 2025; Vestergaard & Schmid, 2026), and acting coherently across government scales (Bärnthaler et al., 2024; Trebeck, 2024; Turner & Wills, 2022) are enabling factors for post-growth government initiatives. In addition, Schmid (2025) and Trebeck (2024) suggest that consistently critiquing the status quo and adopting an openly growth-critical approach could be useful to start challenging the hegemony of pro-growth thinking. Furthermore, some authors have identified participatory democratic processes as useful means to strengthen the inclusiveness of initiatives and the influence of non-corporate actors (Bärnthaler et al., 2024; Turner & Wills, 2022).

Studies of Doughnut economics initiatives found that the accessibility of the visualization, breadth, and agreeability of the approach have been useful to unite a broad base of stakeholders (Cattaneo et al., 2025; Khmara & Kronenberg, 2020; Schmid, 2025), while wellbeing economy studies suggest the breadth could also risk cooptation with pro-growth meanings and policies (Mason & Büchs, 2023; Trebeck, 2024).

## 2.4 Research gap

While there is an emerging body of literature on the characteristics and impacts of government-led post-growth initiatives, to our knowledge, no study has so far examined which barriers and enablers of government-led post-growth initiatives are relevant across governance scales and country contexts. Studies on the wellbeing economy rely either on a small number of cases (Mason & Büchs, 2023; Trebeck, 2024; Turrini et al., 2026) or solely on secondary data (Bärnthaler et al., 2024; Hayden & Dasilva, 2022). Existing research on Doughnut economics initiatives also tend to focus on individual case studies (Cattaneo et al., 2025; Khmara & Kronenberg, 2023; Schmid, 2025; Turner & Wills, 2022; Vestergaard & Schmid, 2026). The present study aims to address this gap by drawing on data from across country contexts and governance scales to identify key barriers and enablers of government-led post-growth initiatives.

The existing literature applies a variety of different frameworks: two wellbeing economy studies explore power dynamics, concentrating on Fuchs' (2016) power framework (Mason & Büchs, 2023) and regulation theory (Bärnthaler et al., 2024). Hayden and Dasilva (2022) discuss their findings in relation to the sufficiency literature while Turrini's (2026) and Trebeck's (2024) studies do not apply an explicit analytical framework. The Doughnut economics studies focus on urban transformations, drawing on ecological economics (Cattaneo et al., 2025), municipalism (Schmid, 2025), urban degrowth economics (Khmara & Kronenberg, 2023), and sustainability governance literature (Turner & Wills, 2022; Vestergaard & Schmid, 2026). To classify barriers and enablers of government-led post-growth initiatives, we adopt what we call a 5Ps framework which captures key factors that the literature has identified so far, including systemic and state-specific factors.

# 3. The 5Ps framework

*Paradigms* refers to the dominant theories, models or discourses that current (economic, social, political, technological, etc.) systems are built around. For instance, the economic growth paradigm manifests materially (e.g. in firms' profit motive, investment strategies, technological development, etc.) and ideologically (e.g. in consumerist understandings of wellbeing, across societal spheres, in political discourses, etc.) (Keyßer et al., 2025; Schmelzer, 2015). *Pressures* refers to crises that indicate problems within existing (economic, social, political, technological, etc.) systems that require transformative responses. Crises tend to affect multiple systems at once which is why we understand them as systemic rather than just political (Buch-Hansen, 2018; Koch, 2020a). For example, the COVID-19 pandemic triggered interlinked economic, political and social crises. However, while crises indicate the need for change, whether or not they facilitate transformative change depends on how actors interpret and respond to them. While being systemic, we understand pressures and paradigms as outcomes of social processes that are influenceable and changeable through collective actions.

To capture and categorize barriers and enablers that relate to state institutions, we adopt the threefold distinction between polity, politics, and policy from political science (Pichler, 2023; Rohe, 1994). We understand these dimensions to be influenced by the systemic factors outlined above. *Polity* refers to the structural features of political institutions, such as their organization and legislation (Pettersson & Sørensen, 2020). *Politics* refers to the processes through which actors pursue competing interests and negotiate change (Patterson et al., 2017; Pichler, 2023). While agency-focused approaches treat power mainly as part of politics (Pichler, 2023), we account also for the structural power inherent in norms and institutions which form the polity. *Policy* includes the instruments and outcomes of political decision-making (Pettersson & Sørensen, 2020; Pichler, 2023). These are shaped by existing institutional structures and actor dynamics but can, in turn, reshape political practice and institutions, either perpetuating unsustainability or driving changes (Pichler, 2023). Thus, despite referring to the outcomes of policymaking processes, we included policy as a dimension of enablers and barriers.

# 4. Methods

Qualitative research allows us to investigate process-oriented questions in depth (Derrington, 2019). We use expert interviews to capture first-hand experiences of people involved with government-led post-growth initiatives who aim to promote processes of change (Meuser & Nagel, 2009). We prepared the sampling process by conducting desk research on the Wellbeing Economy Governments (WEGo), an informal forum of the governments of Wales, Scotland, Finland, Iceland, New Zealand, and Canada dedicated to fostering a wellbeing economy (Mason & Büchs, 2023), and Doughnut economics initiatives in regions and cities (Doughnut Economics Action Lab, 2023). We chose these initiatives as they are the most prominent examples of post-growth government initiatives, due to their transformative potential and their uptake in different localities. While wellbeing economy approaches are mostly adopted at a national and regional level, Doughnut economics initiatives have mostly been adopted by local governments. Thus, studying them together allows us to investigate barriers and enablers that are common across these levels of governance.

To be included in this study's sample, each initiative had to fulfill the following selection criteria:

- The initiative is based on the Doughnut economics or wellbeing economy concept.
- The initiative is based at the national or sub-national governance level.
- The initiative is based in Europe or in a WEGo country.
- The initiative is implemented by the government.

The data included in this study consists of 18 interviews from 14 initiatives across local, regional, and national scales in 11 countries (see Table 1). The initiatives cover all six WEGos (Scotland, Wales, Finland, Iceland, New Zealand, Canada) and eight Doughnut economics initiatives at the city and regional levels. The latter were sampled from Northern Europe, Western Europe, and Southern Europe to include a variety of European regions. However, to ensure the interviewees' anonymity we cannot share the specific countries of the initiatives.

Interviewees included policymakers, policy experts from civil society, and two academics who each had close involvement with at least one of the included initiatives. Speaking to people involved in the initiatives allowed us to access expert knowledge about the processes of the emergence and implementation of these initiatives and related barriers and enablers (Smith & Elger, 2014).

***Table 1***

*Overview of interviews*

| Initiative | # | Interviewee | Governance level | Country context | Code† |
|---|---|---|---|---|---|
| ***Doughnut Economics initiatives*** | | | | | |
| Dougnut 1 | 1 | Policymaker | Local | Western Europe | 1/DE/P/L |
| Doughnut 2 | 4 | Policymaker | Regional | Western Europe | 4/DE/P/R |
| Doughnut 3 | 5 | Expert | Regional | Northern Europe | 5/DE/E/R |
| Doughnut 4 | 7 | Policymaker | City | Western Europe | 7/DE/P/L |
| Doughnut 5 | 8 | Policymaker | Regional | Western Europe | 8/DE/P/R |
| Doughnut 6 | 9 | Academic | Regional | Western Europe | 9/DE/A/R |
| Doughnut 7 | 13 | Policymaker | Local | Southern Europe | 13/DE/P/L |
| Doughnut 8 | 17 | Policymaker | Local | Western Europe | 17/DE/P/L |
| | 18 | Academic | Not applicable | Western Europe | 18/DE/A |
| ***Wellbeing Economy initiatives (WEGos & WEAll)*** | | | | | |
| WEGo 1 | 2 | Policymaker | National | Northern Europe | 2/WE/P/N |
| WEGo 2 | 3 | Policymaker | Regional | Western Europe | 3/WE/P/R |
| WEGo 3 | 6 | Policymaker | Regional | Western Europe | 6/WE/P/R |
| WEGo 4 | 11 | Policymaker | National | Non – European | 11/WE/P/N |
| WEGo 5 | 15 | Policymaker | National | Non – European | 15/WE/P/N |
| WEGo 6 | 16 | Policymaker | National | Northern Europe | 16/WE/P/N |
| WEAll | 10 | Expert | Not applicable | Not applicable | 10/WE/E |
| | 12 | Expert | Not applicable | Not applicable | 12/WE/E |
| | 14 | Expert | Not applicable | Western Europe | 14/WE/E |

*The table provides an overview of the interview sample. It displays the kind of initiative, the role the interviewee held, the governance level and the region the initiative was based in.*

*† The interview code is made up of the following aspects: (1) Interview number. (2) Initiative: DE: Doughnut economics, WE: wellbeing economy (3) Interviewee role: A: Academic, E: Expert, P: Policymaker. (4) Governance level: N: National, R: Regional, L: Local. If the governance level is not applicable, the code does not have a fourth digit.*

Semi-structured interviews were conducted online between November 2023 and March 2024 and lasted between 40 and 80 minutes. The interviews were recorded and transcribed using the intelligent verbatim approach, supported by the automatic transcripts generated by the recording software. Interviews were conducted in English, except for one interview that was conducted in German and translated to English. All transcripts were pseudonymized before analysis, and each interview was given a number and a code identifying the type of initiative.

We analyzed the interviews following a reflexive thematic analysis approach (Braun & Clarke, 2022) with support of the qualitative data analysis software N-Vivo. In line with our overarching research question and based on our 5Ps framework we started by coding the interviews deductively, using the codes of "barriers", "enablers", "paradigms", "pressures", "policy", "politics", and "policy". Throughout the coding process, detailed memos were written to record how barriers and enablers were described and how they relate to our 5Ps framework. After coding the interviews, the notes and the codes were reviewed. Then, in an iterative process of reviewing barriers and enablers from the literature and rereading the coded data segments, enablers and barriers were clustered across systemic and political dimensions.

# 5. Results

Overall, we find that the economic growth paradigm acted as a barrier whereas crises enabled the implementation of Doughnut economics and wellbeing economy initiatives. The three political dimensions related to both barriers and enablers. Figure 1 summarizes the results. Direct quotes are indicated with the interview number and code (see Table 1).

***Figure 1***

*Summary of results: Barriers and enablers for wellbeing economy and Doughnut economics initiatives*

| | BARRIERS | ENABLERS |
|---|---|---|
| **PARADIGMS** | **Growth-dependent economic system**<br>**Dominance of pro-growth discourses** | |
| **PRESSURES** | | **Multiple crises encouraging systems rethinking and urgency to act** |
| **POLITY** | **Siloed organisation of government**<br>**Lack of funding**<br>**Institutional disincentives to change** | **Enabling legislation** |
| **POLITICS** | **Termination of supportive government**<br>**Powerful opposing actors**<br>**Actors' lack of capacity** | **Key individuals and high-level political support**<br>**Supportive alliances**<br>**Participatory governance approaches** |
| **POLICY** | **Breadth of frameworks facilitates cooptation**<br>**Lack of coherence across frameworks** | **Breadth of frameworks allows for broad support base** |

*The two systemic dimensions of paradigms and pressures, as well as the political dimensions of polity, politics, and policy are shown as light grey boxes. The left column lists the barriers, and the*

*right column lists the enablers. A blank space indicates that there was no barrier or enabler identified in the respective dimension.*

## 5.1 Paradigms

### *5.1.1 Barriers*

Several interviewees mentioned the **embeddedness in a growth-dependent economic system** as a barrier. For instance, a representative of a wellbeing economy initiative stated that this embeddedness limits "*the capacity for any single country to ultimately transform or deal with, let's say, the climate crisis"* (Interview 12, WE/E). Interviewees at the regional and local scale underscored the limited ability of subnational scales to move beyond framework conditions set by the national level because *"at the moment, the growth idea, economic growth, is so embedded in kind of everything"* (Interview 5, DE/E/R). One interviewee stated that within their institution economic growth was seen as *"more important than nature"* (Interview 16, WE/P/N) because it was understood as necessary to achieve social outcomes.

Furthermore, interviewees described various expressions of hegemonic **pro-growth discourses**. For instance, individuals sympathetic to post-growth working in administrations reported that they are confronted with attitudes by colleagues which imply that "*sustainability, eco [...] that's all well and good, but the numbers, that's what counts"* (Interview 7, DE/P/L). Having conversations in which the pursuit of economic growth would be questioned was described as highly polarizing, because *"it scares so [many] people to talk directly on that question"* (Interview 13, DE/P/L) and *"if you say that explicitly, you are completely shutting off the conversation with your conversation partner"* (Interview 18, WE/A).

### *5.1.2 Enablers*

We did not identify any enablers in the dimension of paradigms.

## 5.2 Pressures

### *5.2.1 Barriers*

We did not identify any barriers in the dimension of pressures.

### *5.2.2 Enablers*

Interviewees described the exposure to **multiple crises**, for instance the COVID-19 pandemic, as placing wellbeing higher on the public and government agenda and encouraging system rethinking. For example, interviewee 1 (DE/P/L) remembered how "*some exceptional actions"* related to the energy crisis in the wake of the war in Ukraine, as well as the "*double context of accelerating alerts on global change etc. made it really important for the city to really take action and to decide to move forward within [the Doughnut economics] framework"*.

Moreover, interviewees described how multiple crises led to a shift in narratives and a perceived urgency to act. People's experience of multiple crises *"just makes it easier and* *easier obviously to tell the story that the system is not working, our economy is not working because everybody can see it now"* (Interview 14, WE/E). Interviewee 20 (WE/E) reported that for one WEGo *"[...] COVID and that increasing public dissatisfaction, that was a big impetus and catalyst [...]"*.

On an international level, the 2007/2008 financial crises prompted the emergence of a discourse around alternative measures of welfare which multiple interviewees mentioned, referring to the Stiglitz-Sen-Fitoussi report (Stiglitz et al., 2009). At the same time, the introduction of initiatives influenced political norms within institutions: *"you can see the change internally in the type of conversations that colleagues have with each other, the types of considerations that get raised and then the tone and the focus of the papers that get produced and the balance that's struck between competing objectives"* (Interview 15, WE/P/N).

## 5.3 Polity

### *5.3.1 Barriers*

Across interviews, the **siloed organization of governmental institutions** was described as hampering integrative, post-growth-oriented policymaking. Interviewee 14 (WE/E) stated, *"even though the wellbeing economy is meant to be this overarching thing, that's about aligning and designing the whole of government, they've effectively managed to silo it as well."* Other interviewees referred to silos related to thinking about environmental and social challenges separately and limited responsibilities of people working on the wellbeing economy.

In addition, a **lack of funding** led to the foreseeable end of multiple post-growth initiatives. This was explicitly compared to other departments within government that were more aligned with economic priorities: *"for example, in the Ministry of Finance, the thinking is quite old school and [...] they have the money, so they have the power*" (Interview 16, WE/P/N).

**Institutional disincentives to change** translate into change-averse attitudes of individual policymakers which led one interviewee to remark that *"a lot of people are just very resistant to change in general"* (Interview 17, DE/P/L). Another interviewee observed that *"really changing the way organizations act is quite a cumbersome process and I think one of the big challenges that we have is, is the attitude to risk. [...] a lot of politicians, but also a lot of civil servants, don't want [...] to try something new that might go wrong because [...] they're worried they get criticized for it"* (Interview 14, WE/E).

### *5.3.2 Enablers*

Passing **enabling legislation** that provides a framework for governments to prioritize human wellbeing and environmental outcomes was suggested by two interviewees as a useful step to enforce post-growth approaches long-term. Interviewee 3 (WE/P/R) shared the experience that in their constituency the *"big and enabling legislative framework [...] is dictating everything we do and everything we want to achieve in the next years, decades and* *years to come."* Interviewee 10 (WE/E) agreed that *"legal frameworks and change is really important, like rights of nature and ecocide law to kind of say, hey look, these are the boundaries within which every government needs to operate."*

## 5.4 Politics

### *5.4.1 Barriers*

Existing government-led post-growth initiatives tend to be led by specific governing parties or coalitions and are thus vulnerable to **changes in or termination of a supportive government.** In several cases, post-growth initiatives were severely weakened or even abandoned after a supportive government ended or a supportive policymaker lost power or stepped down. Being supported by a specific political party or politician can also generate more general tensions, e.g. as reflected on by interviewee 4 (PG/P/R): "*So [...] on one hand, it [support from a governing party] has helped us to accelerate and to give us the possibility via subsidies to start the work. But on the other hand, it's also a bottleneck [...] because it doesn't allow us to go broader"*. Interviewees described that the political cycles lead to short-term thinking from politicians who *"want to see short term results"* (Interview 3, WE/P/R).

**Powerful opposing actors** were identified as a key barrier for initiatives. These were mainly economic actors, including *"multinationals and big organizations"* (Interview 10, WE/E), *"some companies"* (Interview 1, DE/P/L), and *"global finance"* (Interview 12, WE/E). Interviewee 5 felt that more generally *"people in power they do not want things to change because otherwise they would have to struggle for their power"* (DE/E/R).

Interviewees described a **lack of capacity** of policymakers to "*be burdened with something on top of their daily task"* (Interview 7, DE/P/L). One interviewee from a Doughnut initiative anticipated that "*if we were to implement the tool I mentioned [...] people were going 'ohh well, that will be so much work', and 'don't we do a lot of good things already' and 'why do we need to use this'"* (Interview 17, DE/P/L).

### *5.4.2 Enablers*

Initiatives often depend on **key individuals** who are intrinsically motivated to drive the initiative, as well as on **high-level political support**. An interviewee from a WEGo observed that *"anywhere around the world where this has taken hold there usually is a president or a prime minister or a finance ministry or treasury, or a king who makes this their thing that they push"* (Interview 11, WE/P/N). This enabler directly relates to the reported barrier of vulnerability to changes in a supportive government or policymaker. Interviewee 17 (DE/P/L) described that they *"feel it's very dependent on people. So if you have people in a department that [...] go like 'we need to do something with this, I'm gonna set this up [...]"*.

**Supportive alliances** between peer initiatives (e.g. through the WEGos or Doughnut Economics Action Lab), as well as support from international organizations (e.g. the Organization for Economic Co-operation and Development), were considered useful for peer learning and capacity building. While civil society action as a driving force was rather the exception, Interviewee 12 (WE/E) described how in one WEGo initiative *"it was really like a mid-level civil servant who [...] sort of rallied and developed a really strategic group of citizens and advocacy organizations to put pressure on the government to ultimately pass this type of [WE] legislation"*.

Having a wide variety of societal actors take part was often considered a necessity for an initiative to be successful. For example, several initiatives drew on **participatory governance approaches** to shape government strategies. These approaches were discussed as theoretically having the potential to contribute to redistributing power to citizens, raising public support, and providing legitimacy for the initiative with the public. However, interviewees provided a mixed picture as to whether the recommendations of respective bodies were being considered in decision-making. Moreover, interviewees criticized problems of accessibility and representation in participatory processes.

## 5.5 Policy

### *5.5.1 Barriers*

The **breadth of the concepts facilitates cooptation** and leads to a disconnection between the established tools and the actual priorities in policymaking. The lack of a common definition of the core concepts within government has allowed for a stretch of the wellbeing economy and Doughnut economics concepts towards pre-established, economic growth–focused policy goals. As Interviewee 10 (WE/E) put it: *"I feel like a lot of the wellbeing frameworks that we're seeing are not really conceptual frameworks. They're like selections of indicators [...] but nothing in them really says that our economies are embedded within our societies and our environments. And I think that's a huge paradigm shift that we're still missing"*.

In addition, there is a **lack of coherence**, e.g. between different sustainability visions and indicator frameworks or between different government departments. First, this refers to the multiplicity of sustainability visions and related indicator frameworks used in policymaking within the same government. For example, Interviewee 16 (WE/P/N) reported that "*people don't understand [...] why don't we talk about the [...] Sustainable Development Goals or why don't we talk about green transition? [...] And the problem is that there are too many concepts like in a competition."* Second, several interviewees also mentioned a lack of coherence between different government departments due to siloed ways of working within government. For example, Interviewee 13 (DE/P/L) suggested that the Doughnut portrait developed in the environmental department of the city would not lead to "*an economic strategy in that sense also because the city council, parallel to this developed [its] own economic strategy. And both things don't talk too much [to each other]."*

### *5.5.2 Enablers*

The **breadth and agreeability** of the concepts of the wellbeing economy and Doughnut economics **allowed for a broad support base**. Interviewees reported that they would start their work from easily agreeable points and frame their work in a non-confrontational way: *"quite a lot of people can sign up to [the wellbeing economy] I think, which is the strength and the weakness at the same time, and partially deliberate. But then also you know [this]* *opens it up to [...] being co-opted by the government and by other people with different kind of meanings*" (Interview 14, WE/E).

# 6. Discussion

## 6.1 Main findings

Overall, our findings strengthen the evidence base on barriers and enablers for government-led post-growth initiatives, as they confirm that many factors that have previously been identified for individual or a smaller number of cases apply across different contexts. We find that wellbeing economy and Doughnut economics initiatives remain very restricted in their scope of action by the dominant economic growth *paradigm*. The limiting effect of the growth paradigm is reflected in the initiatives' dependence on growth-based funding, the hegemony of the pro-growth discourse that makes it inopportune to criticize growth or openly advocate post-growth, and the continued importance of GDP as an indicator in policymaking.

While previous empirical studies on beyond GDP initiatives find *crises* to play an ambiguous role, in our study interviewees described them as enablers, in line with the conceptual literature on post-growth transformations (Buch-Hansen, 2018; Koch, 2020a). In addition, in our study highly motivated individuals and high-level support from within political institutions have facilitated several government-led post-growth initiatives, supported by interactions with civil society. However, therefore the transformative capacity of government-led post-growth initiatives is vulnerable to changes in political power constellations. Moreover, policymakers who aim to foster growth-critical perspectives face tensions between appealing to a broad stakeholder base and ensuring the longevity of the project, on the one hand, and aiming to advocate for bold changes and avoid cooptation on the other.

## 6.2 Contributions and limitations

Our research empirically contributes to understanding the role of government-led post-growth initiatives in post-growth transformations in two ways. First, while previous research is based on a small number of empirical cases of either wellbeing economy or Doughnut economics initiatives, we collected primary data from wellbeing economy and Doughnut economics initiatives across governance levels in nine European countries, plus New Zealand and Canada. Second, we adopt a 5Ps framework to categorize barriers and enablers across the systemic dimensions of paradigms and pressures, and the political dimensions of polity, politics and policy.

Our study faces several limitations. As we draw on insights from a range of countries and governance scales, we cannot provide in-depth analysis of the institutional settings, actor constellations, or power dynamics in respective contexts and governance scales. Moreover, this study does not provide an empirical analysis of the relative importance of different barriers and enablers.

## 6.3 Implications and future research

Our findings suggest that the implementation of post-growth government initiatives is shaped not only by political dimensions but also by paradigms and pressures that enable or restrict transformative change. The following section discusses the implications of our findings for research and policymaking and outlines priorities for future research.

By their very nature, governmental institutions favor dominant ideas and therefore tend to have a conserving function (Bell, 2011; Buch-Hansen, 2018; Jessop, 2016). Nevertheless, in line with our theoretical understanding of the state as being co-constituted by civil society (D'Alisa & Kallis, 2020; Koch, 2022b), and as identified in previous empirical studies, our results suggest that strategic alliance between post-growth-minded policymakers and civil society actors could be a promising avenue to challenge institutionalized growth-based logics of policymaking and to put pressure on governments (Trebeck, 2024; Turrini et al., 2026). In addition, it could contribute to the necessary cultural changes by changing mentalities and creating consent for post-growth transformations in both political and civil society (Bärnthaler, 2024b; D'Alisa & Kallis, 2020).

Developing a critical social consciousness of the current growth hegemony could be the starting point to transform the state through its imperative for democratic legitimation as suggested by Hammond (2025). As some interviewees described, the unfolding of multiple crises may accelerate processes of new "meaning making" (Hammond, 2025; Koch, 2020a). At the time of the interviews, interviewees noted a growing awareness of multiple crises and an increased focus on wellbeing as a policy goal, both within governments and civil society. Some initiatives in our study influenced the discussions and priority setting within their institutions, providing examples of practical, post-growth aligned changes in the values of policymakers directly involved in the initiatives (D'Alisa & Kallis, 2020; Wallace, 2019). This suggests that there could be potential for post-growth discourses to gain traction. Framings emphasizing wellbeing and equality increases in post-growth futures (Paulson & Büchs, 2022) and showing socially desirable benefits of post-growth policies (Strunz & Schindler, 2018) could resonate with a wide range of people in society. Gaining discursive power could be a first entry point for post-growth initiatives to strengthen their position (Fuchs et al., 2016).

Policymakers aiming to introduce growth-critical perspectives face tensions between broadening stakeholder support and maintaining project longevity, on the one hand, and advocating bold changes and avoiding cooptation on the other. Our findings support Schmid's (2025) observation that policymakers in Doughnut economics initiatives strategically avoid direct critiques of economic growth to maintain broader appeal. However, avoiding direct critique may leave initiatives vulnerable to green growth co-optation. Thus, we would underscore the importance of a continuous critique of the status quo, which has been pointed out in previous studies on post-growth government initiatives as well. To support policymakers in introducing a growth-critical discourse, Schmid (2025) proposes a five-step strategy: (1) clarifying what is meant by growth (disambiguation); (2) examining how prioritization of growth currently informs policymaking (scrutinization); (3) acknowledging where prioritizing growth contradicts social and environmental objectives (acknowledgement); (4) identifying areas where growth orientation is counter-productive and exploring alternative approaches (partitioning); and (5) exploring (municipal) growth-dependencies and alternative strategies (generalization). This stepwise approach may enable more constructive debates about the challenges of the continuous pursuit of economic growth.

To date, much post-growth research has focused on identifying policies for post-growth societies (e.g. Cosme et al., 2017; Fitzpatrick et al., 2022; Kallis et al., 2025). However, our results suggest that policies have limited transformative potential in the given paradigms, polity and politics. Thus, we argue that identifying the systemic and political conditions under which transformative changes could be implemented would be more effective in fostering transformations. That said, if incremental policy changes were chosen strategically to reflect degrowth-aligned values, they could contribute to cultural changes, which would allow for more ambitious policy changes later on (D'Alisa & Kallis, 2020). For example, universal basic services could provide a first-hand "experience" of alternative provisioning to the majority of people, appealing to high-income and low-income groups alike as well as to people with a variety of values (Bärnthaler, 2024a; Paulson & Büchs, 2022).

Future research could explore which strategies agents of change in political institutions take to promote post-growth approaches in policymaking, as our findings underscore the influential role of key individuals in initiating initiatives within government institutions. In addition, future case studies of government-led post-growth initiatives could investigate power relations and possible alliances that enable or restrain post-growth policymaking in specific contexts and provide an empirical analysis of the relative importance of different barriers and enablers. Moreover, our findings could inform further theoretical contributions on the role of the state in post-growth transformations.

# 7. Conclusion

In this study, we have asked the question: which barriers and enablers of post-growth government initiatives are common across wellbeing economy and Doughnut economics initiatives, country contexts, and governance levels? To answer this question, we collected primary data from 18 interviews across local, regional, and national scales from 11 countries. We applied a 5Ps framework, capturing paradigms, pressures, polity, politics, and policy to categorize barriers and enablers.

Overall, we find that the hegemony of the economic growth paradigm limits the scope of action of wellbeing economy and Doughnut economics initiatives and shapes political dimensions as reflected in the dominance of growth-focused government departments, funding, policy priorities and political actors. Highly motivated individuals and high-level support from within political institutions have facilitated several government-led post-growth initiatives, supported by interactions with civil society. Policymakers who aim to foster growth-critical perspectives face tensions between appealing to a broad stakeholder base and ensuring the longevity of the project, on the one hand, and aiming to advocate for bold changes and avoid cooptation on the other. Moreover, despite the focus on policies within the post-growth literature, our results suggest that they have limited transformative potential in the current system.

Structural changes are required to address one of the key barriers to transformation: the economic growth paradigm. For instance, developing civil society initiatives that can directly deliberate and develop post-growth approaches with citizens as well as a greater presence of post-growth approaches in the media and economics education would be needed for post-growth discourses to become more accepted, and to facilitate understandings of how growth dependencies can be addressed. Wellbeing economy and Doughnut economics initiatives themselves could aim to openly criticize the continuous pursuit of economic growth and collaborate with civil society initiatives that support post-growth-aligned mindsets among citizens. At the same time, initiatives could strive to enable deliberative forms of democratic participation, to open spaces to debate and challenge the hegemony of economic growth within institutions in order to allow for more ambitious post-growth policies in the long term.

# Acknowledgments

We would like to thank all interviewees who gave their time generously to participate in this study. Ethics approval was obtained from the University of Barcelona (Bioethics Commission), and from the University of Leeds (Faculties of Business, Environment, and Social Sciences Ethics Committee, reference 0937), covering full informed consent from interview participants and anonymization of interview data in research outputs.

This research received funding from the European Union's Horizon Europe research and innovation programme under grant agreement number 101094211 (ToBe: "Towards a Sustainable Wellbeing Economy: Integrated Policies and Transformative Indicators") and UK Research and Innovation (UKRI) under grant agreement number 10052343. Views and opinions expressed are, however, those of the authors only and do not necessarily reflect those of the European Union or UKRI. Neither the European Union, the granting authority, nor UKRI can be held responsible for them.

# Data availability statement

The data set underlying this study is not currently publicly available to protect the anonymity of research participants.

# Author contributions

**Laura Angresius:** Conceptualization, Methodology, Investigation (data collection), Formal analysis, Writing – Original Draft, Writing – Review & Editing; **Milena Büchs:** Conceptualization, Methodology, Investigation (data collection), Writing – Review & Editing, Supervision, Funding acquisition; **Alessia Greselin:** Investigation (data collection), Writing

– Review & Editing; **Daniel W. O'Neill:** Writing – Review & Editing, Supervision, Funding acquisition.

## Disclosure statement

The authors report there are no competing interests to declare.